\documentclass[namedreferences]{solarphysics}

\usepackage{amsmath}
\usepackage[hyperref,optionalrh,solaromanenum]{spr-sola-addons} 
\usepackage{color}                       
\usepackage{breakurl}                         
\usepackage{graphicx}  
\usepackage{epstopdf} 
\usepackage{gensymb}
\usepackage{float}
\usepackage{caption}
\usepackage{ifpdf}

\begin{document}
\begin{article}
\begin{opening}

\title{The Effect of Limited Sample Sizes on the Accuracy of the Estimated Scaling Parameter for Power-Law-Distributed Solar Data}

\author[addressref={aff1,aff2}, corref, email={elke.dhuys@observatory.be}]{\inits{E.}\fnm{Elke}~\lnm{D'Huys}}
\author[addressref={aff1}]{\inits{D.}\fnm{David}~\lnm{ Berghmans}}
\author[addressref={aff1,aff3,aff4}]{\inits{D.B.}\fnm{Daniel B.}~\lnm{Seaton}}
\author[addressref={aff2}]{\inits{S.}\fnm{Stefaan}~\lnm{Poedts}}
\address[id=aff1]{Royal Observatory of Belgium,  Ringlaan -3- Av. Circulaire, 1180 Brussels, Belgium}
\address[id=aff2]{Katholieke Universiteit Leuven, Centre for mathematical Plasma-Astrophysics, Ce\-les\-tij\-nen\-laan 200b - bus 2400, 3001 Leuven, Belgium}
\address[id=aff3]{Cooperative Institute for Research in Environmental Sciences, University of Colorado, Boulder, Colorado U.S.A.}
\address[id=aff4]{NOAA National Centers for Environmental Information, Boulder, Colorado U.S.A.}

\begin{abstract}
Many natural processes exhibit power-law behavior. The power-law exponent is linked to the underlying physical process and therefore its precise value is of interest. With respect to the energy content of nanoflares, for example, a power-law exponent steeper than 2 is believed to be a necessary condition to solve the enigmatic coronal heating problem. Studying power-law distributions over several orders of magnitudes requires sufficient data and appropriate methodology. In this paper we demonstrate the shortcomings of some popular methods in solar physics that are applied to data of typical sample sizes. We use synthetic data to study the effect of the sample size on the performance of different estimation methods and show that vast amounts of data are needed to obtain a reliable result with graphical methods (where the power-law exponent is estimated by a linear fit on a log-transformed histogram of the data). We revisit published results on power laws for the angular width of solar coronal mass ejections and the radiative losses of nanoflares. We demonstrate the benefits of the maximum likelihood estimator and advocate its use.
\end{abstract}
\end{opening}

\section{Introduction}

Power laws are ubiquitous in solar physics and have been found in studies of the flux of solar flares (in soft and hard X-rays, EUV, and radio wavelengths), flare waiting times (the time between consecutive flares), abundance enhancements, and waiting times for solar energetic particle events; and in solar wind measurements. A comprehensive overview is given by \cite{Aschwanden2014}. 

For the purpose of this paper, we define a power-law function as follows:
\begin{equation}
f(x) \sim x^{- \alpha},
\end{equation}
where $\alpha$ is called the exponent, scaling factor, or scaling parameter. In solar physics, this scaling parameter typically lies within the range $ 1.1 \leqslant \alpha \leqslant 3.0$ \citep{Aschwanden2014}. Often only the tail of a distribution can be described by a power law. In this case, we define $x_{\rm min}$ as the minimal value for which the power-law behavior is observed. Similarly, we take $x_{\rm max}$ as the largest $x$-value for which the power-law properties hold. 

Power laws are important because they suggest scale invariance: small and large events have the same properties because they are initiated by the same underlying physical process. For instance, a strong correspondence is found between the power-law behavior for the X-ray radiation received from solar microflares and from major flares --- even though these events are observed with separate instruments and on widely different scales --- and indeed solar physicists believe that all flares are triggered by one and the same process, called magnetic reconnection.

The power-law distributions observed for different solar parameters are often attributed to self-organized criticality \citep[SOC;][]{Bak1987}. SOC describes how in a minimally-stable dynamical system a minor event can start a chain reaction by which any number of elements in the system may be affected. Applying this theory, \cite{Lu1991} argued that a solar flare can be interpreted as an avalanche of many small reconnection events, resulting in a power-law distribution for the flare occurrence. Other physical processes are invoked to explain the presence of power laws in solar physics as well. In the case of solar flare waiting times, for example, \cite{Boffetta1999} claim that SOC cannot reproduce the observed power-law behavior, while MHD turbulence models can.

The presence of a power law --- and the deviations from it --- provides important information about the underlying physics of the measured events. Parameters derived from observational power laws are compared to theoretical predictions in order to validate the theory. The importance of correctly estimating the exponent is illustrated by the argument made by \cite{Hudson1991} that the solar flare energy distribution must have a power-law index $\alpha > 2$ in order for nanoflares to explain coronal heating. A power law this steep implies that a large number of small events with low energies occurring simultaneously would supply sufficient energy to heat the corona. Therefore, many authors have spent considerable effort trying to determine the exact scaling parameter for solar flares, resulting in a wide range of estimations of $\alpha$, which left the question undecided.

Unfortunately, many authors apparently underestimate the care that must be taken when attempting to estimate power-law scaling parameters derived from data. With the vast increase in observational data volume, robust statistical analysis of large distributions is becoming an increasingly important issue in solar physics, and we will show in this paper that not all authors have realized this fully when characterizing the power-law dependence of their data. While the application of inappropriate fitting techniques to power-law-dependent data may not necessarily completely undermine the analysis, such analyses can induce error in the estimated power-law parameters. 

There are many recent reports in the literature about the power-law nature of various solar phenomena, but not all of these authors report using appropriate power-law estimation techniques, and many of the conclusions presented in these papers would benefit from a re-analysis using the methods discussed in this paper. Among them are analyses of the properties of solar flares associated with and not associated with coronal mass ejections \citep{Yashiro2006}, analyses of the rate of active region transient brightenings observed in soft X-rays \citep{Shimizu1995} and quiet-Sun transient brightenings observed in extreme-ultraviolet \citep{Berghmans1998}. Even our own analysis in \cite{DHuys2014} used a graphical method (a linear fit on a log-transformed histogram of the data) to fit a power law to the size of coronal mass ejections with various properties. Although the particular fitting technique these authors used is certainly not grounds to invalidate the conclusions of any of these papers, a re-analysis with more appropriate techniques would certainly improve the support for these papers' results. Indeed, we will examine several of the conclusions by the latter authors in this paper.	

We will discuss different power-law estimation methods in more detail below. We focus on the effect of the sample size on the accuracy of the various estimation methods and apply them to observational solar data. 

\section{Estimation Methods for the Power-Law Exponent}

Authors can apply various methods to estimate the scaling parameter $\alpha$ for their power-law-distributed data. Here we discuss the most common approaches to estimate the exponent and highlight their strengths and shortcomings. 

It is important to note that in this paper we only focus on methods to estimate this scaling parameter and assume that the data are, in fact, power-law-distributed, which may not always be the case. In reality, it is often quite difficult to determine whether a power law is an appropriate distribution to describe observational data. For example, log-normal or exponential behavior is hard to distinguish from a power law, especially when only a limited range of the distribution is investigated. Additionally, turn-overs are often observed, beyond which the power-law behavior no longer holds. One should therefore carefully consider whether a power law is in fact an appropriate model for the data at hand. This aspect, however crucial, is beyond the scope of this paper. \citet{Clauset2007} describe how, for example, a likelihood ratio test can be used to determine whether other models fit the data better.

In this work, we focus on the effect of the sample size on different estimation methods for the power-law exponent. To test these methods independently from any other effect, we use perfectly power-law-distributed sample data, while at the same time acknowledging that such perfect data are hardly ever found in real life. 

\subsection{Graphical Methods}

In search of a power law, authors traditionally create a log-transformed histogram of their data. The power law then manifests itself as a straight line of data points in this graph. Frequently, authors apply a least-squares linear fit to estimate the slope of this line, which corresponds to the power-law scaling parameter. Bins without observations are not taken into account (because $\log(0)$ is undefined) and in some cases bins with low counts are excluded as well \citep{White2008}. This technique is widely used and straightforward to apply, but is prone to errors. Most importantly, this method is very sensitive to the choice of bin width and the sample size. This is not surprising, as binning results in a loss of information about the distribution of the data points within a bin \citep{Clauset2007}. 

Improvements to this method can be made by using logarithmic binning: a larger (linear) bin size for the less dense populated tail of the distribution ensures more data points per bin and thus this approach reduces the large statistical errors that are observed in the case of linear binning  where many bins are scarcely populated. It is important however to know that logarithmic binning will provide a slope value of $-\alpha + 1$, not $-\alpha$. This has often been overlooked in previous studies and can be avoided by using normalized logarithmic binning \citep{White2008}. 

Interestingly, \cite{Goldstein2004} also claim that in the case that only the first five points in the frequency distribution on a log-log scale are used for the estimation of the slope, the true scaling parameter is often quite well reproduced, albeit with a very large variance. Indeed, these first five bins contain most of the data and thus we may obtain an accurate estimate of the slope with this limited amount of data, while at the same time avoiding the counting problems that often occur in the tail of the distribution in the case of small sample sizes.

\subsection{Maximum Likelihood Estimation}

The method of maximum likelihood provides an accurate estimate of the scaling parameter $\alpha$ through a simple calculation. If we assume that our data follow a power-law distribution for $x \geqslant x_{\rm min}$, the \textit{maximum likelihood estimator} (MLE) for continuous data is given by
\begin{equation}
\hat{\alpha} = 1 + n \left [ \sum\limits_{i=1}^n \ln{\frac{x_i}{x_{\rm min}}} \right ]^{-1}
\label{MLEformula}
\end{equation}
where $x_i,\ i = 1 \dots n$, are the observed values of $x$ such that $x_i \geqslant x_{\rm min}$  \citep{Muniruzzaman1957, Clauset2007}. The standard error on $\hat{\alpha} $ is then
\begin{equation}
\sigma = \frac{\hat{\alpha} - 1}{\sqrt{n}} + \mathcal{O}(1/n).
\label{MLEerrorformula}
\end{equation}

\cite{Goldstein2004} showed that the variance of the estimates obtained with MLE is notably lower than that of the estimates using a linear fit on the first five bins in the frequency distribution. In fact, MLE has mathematically been shown to be the minimum variance unbiased estimator \citep{White2008}, that is, MLE has a lower variance than any other unbiased estimator for all possible values of the scaling parameter. This makes MLE the most accurate and robust method to estimate the power-law scaling parameter $\alpha$. In addition, this method produces a valid confidence interval for $\alpha$, which is not the case for other approaches. 

However, there are drawbacks to the MLE technique. As we will discuss, MLE is sensitive to small sample sizes. Additionally, the result is heavily dependent on the choice of the parameter $x_{\rm min}$. Taking $x_{\rm min}$ too low is especially problematic because it results in the inclusion of non-power-law-distributed data points in the calculation, which, in turn, induces erroneous results. A value for $x_{\rm min}$ that is chosen a bit too high is less of a problem: in this case the MLE estimate remains a good approximation for $\alpha$, even if valid data points are ignored during the calculation \citep{Clauset2007}. 

A cut-off at the higher end of the $x$-axis, beyond which the data are no longer power-law-distributed, is often observed and can be handled using MLE; however, this is not straightforward. \cite{White2008} provide an MLE formula for a truncated power-law distribution. Note that to solve this equation numerical methods are required. Despite these complications, deviations from the power-law behavior should not be ignored, especially because they may prove to be vital to understanding the underlying physical process \citep{White2008}. 

Due to the inaccuracy of visual methods, $x_{\rm min}$ should not be derived from the logarithmically transformed histogram. Instead, \citet{Clauset2007} advocate the more objective and statistically sound method of choosing the value of $x_{\rm min}$ that reduces the difference between the probability distribution of the real data and that of the best fitting power law, measured using a Kolmogorov-Smirnov (KS) statistic. The preferred value for $x_{\rm min}$ is the one that minimizes this KS statistic (see, for example, Section~\ref{stealthSection}).

\subsection{Alternative Methods}
Other strategies exist to estimate the scaling parameter for power-law-distributed data. As an alternative graphical method one may use the cumulative distribution function (CDF). Linear regression can be applied to the log-transformed CDF and the resulting slope corresponds to $\alpha - 1$. The main advantages of using the CDF are that it is straightforward to calculate and that no binning is required, which in turn avoids a bias caused by the choice of bin width or the presence of empty bins \citep{White2008}. In addition, these authors claim that the CDF is more robust to fluctuations in sample size than other graphical methods. Nevertheless, the MLE method still outperforms this technique in both accuracy and precision.

A Bayesian approach is another way to estimate the scaling parameter $\alpha$ of power-law-distributed data. As \cite{Wheatland2004} explains, estimating a power-law exponent means finding a probability function for this parameter and taking the maximum of this function as the most likely value for the exponent, while the width of the probability function is a measure for the uncertainty of this estimate. In fact, MLE is a special case of the more general Bayesian approach, with the assumption of a uniform prior \citep{Wheatland2004}.

\section{Sample Size Influence}
To illustrate the influence of the sample size on the estimation of $\alpha$, we created artificial datasets of various sizes and applied some of the techniques described above. It is easily understood that in case of small samples, the histogram bins will be too sparsely populated to be well-fit by the graphical methods. \citet{Clauset2007} proposed a sample size $n \geqslant 50$ as a rule of thumb to obtain a reliable estimate with the MLE method. 

We created the random samples for this study using the technique described by \citet[Appendix D]{Clauset2007} and \cite{Newman2005}. Given an array $r$ of uniformly distributed random numbers in the range $0 \leqslant r < 1$, we calculated a random sample $x$ that is power-law-distributed as follows: 
\begin{equation}
x = x_{\rm min} (1- r)^{-1/(\alpha-1)}.
\end{equation}
We have arbitrarily set $x_{\rm min} = 10$. We used values between $1.1$ and $3.0$ for $\alpha$ and evaluated the performance of the different estimation methods as a function of sample size and this scaling parameter.

We can directly apply the MLE method to these random samples of power-law-distributed data, using Equations~\eqref{MLEformula} and~\eqref{MLEerrorformula} above. For the graphical methods, we first need to create a histogram. The artificial power-law distribution described above converges to zero for large $x$-values. Although unlikely, it is possible that a given sample will contain a very large $x$-value, which would lead to a huge number of empty bins in its histogram, creating memory issues during the calculations. To avoid this in the simplest way possible, we have cut the histogram at the higher end (we arbitrarily chose $x_{\rm max} = 1000$) -- although storing the indices and the numbers only for the bins with non-zero counts could have been an alternative approach. This results in lower actual sample sizes because some of the data points are ignored in the calculation. For ease of comparison between different methods we plot the sample size for the complete sample (including $x_i \geqslant 1000$) in all figures throughout this paper.

\subsection{Graphical Methods}

To apply the graphical method to our samples, we performed the log-log transform of the histogram of the data for each bin with more than one data point. We then calculated the slope of the resulting line using a least-squares linear regression technique to obtain an estimate for the scaling parameter. As a possible improvement to this method, we repeated this calculation using only the first five bins of the histogram that contained more than one element. 
\ifpdf
\begin{figure}        
   \centerline{\hspace*{0.015\textwidth}
               \includegraphics[width=0.515\textwidth,clip=]{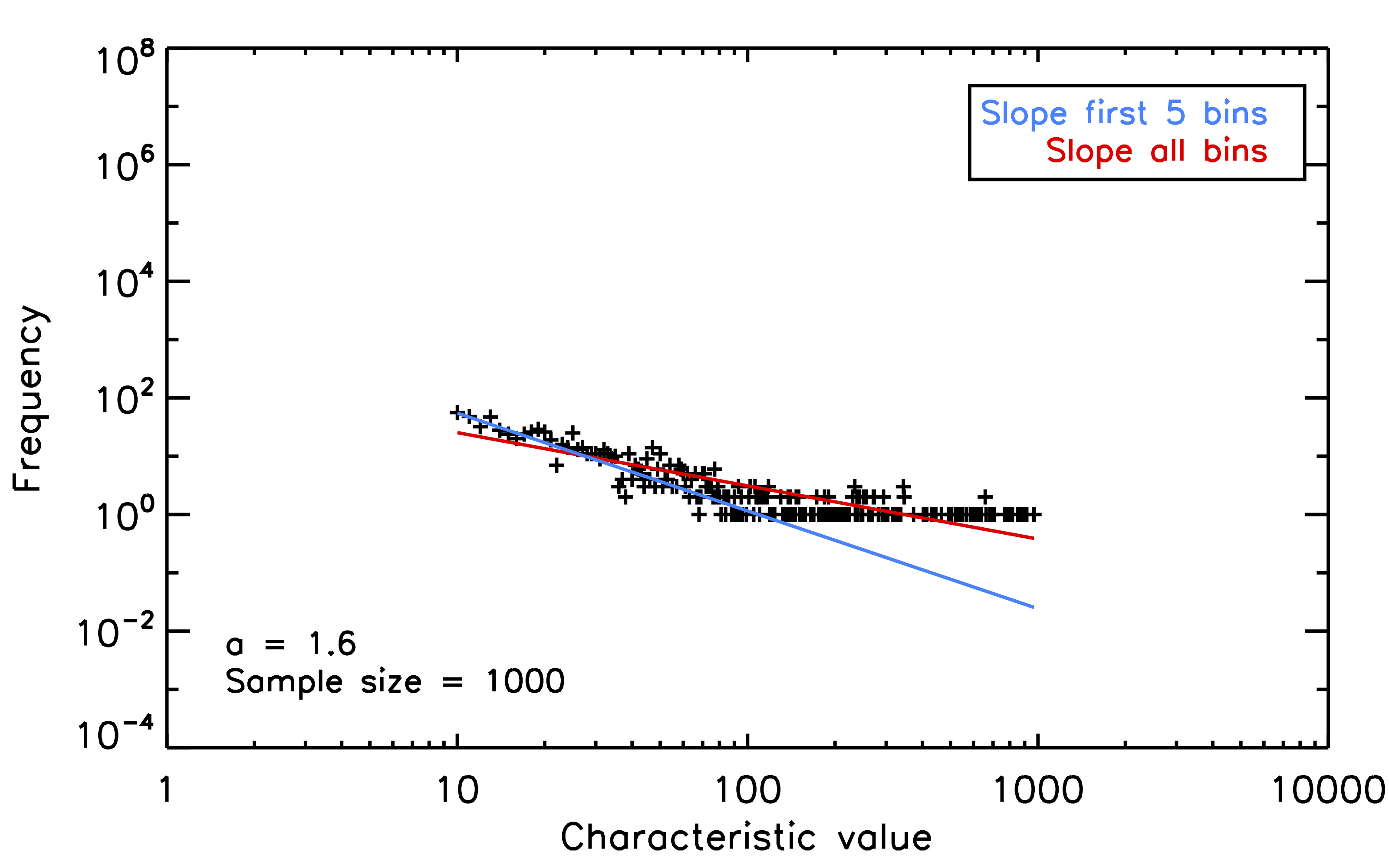}
               \hspace*{-0.03\textwidth}
               \includegraphics[width=0.515\textwidth,clip=]{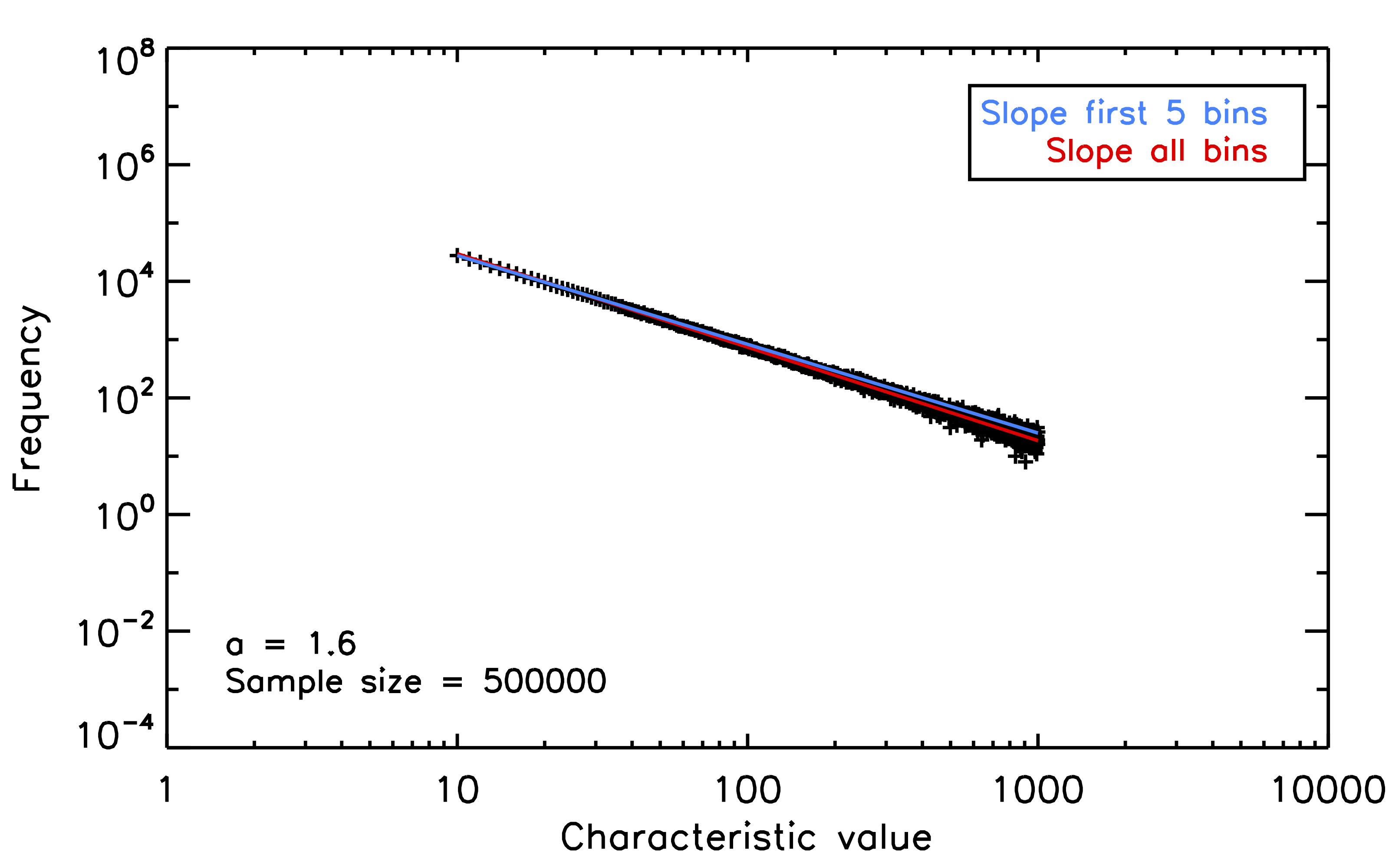}      
               }
\caption{Log-transformed histograms for random samples with 1000 (left) and 500~000 (right) elements that were drawn from a power-law distribution with scaling parameter $\alpha = 1.6$. The red and blue lines are the best linear fits based on all bins and on the first five bins with more than one data point, respectively. The fitted slope values are presented in Figure~\ref{samplesizeinfluence_allbins_5bins}.}
\label{samplesizehistograms}
\end{figure}

In Figure~\ref{samplesizehistograms} we show the log-transformed histograms for a small and a large power-law-distributed sample. These are random samples with scaling parameter $\alpha = 1.6$. The red lines indicate the best fit to the data using a linear regression technique on all histogram bins with more than one data point. The blue line indicates the best fit using only the first five such bins. It is immediately clear that for a small sample size the estimation of the slope is difficult due to under-sampling: the bins for the lower values are well filled, but the higher ones are not. This heavy tail has a strong influence on the slope estimation. Only when the sample size is sufficiently large, can we resolve this problem of under-sampling in the higher bins and obtain a reliable estimate of $\alpha$, as is shown in the right panel of Figure~\ref{samplesizehistograms}. A supplementary movie shows the improvement of the estimates for $\alpha$ when applying the graphical methods to samples of increasing size. 

\begin{figure}        
\centerline{\includegraphics[width=0.7\textwidth]{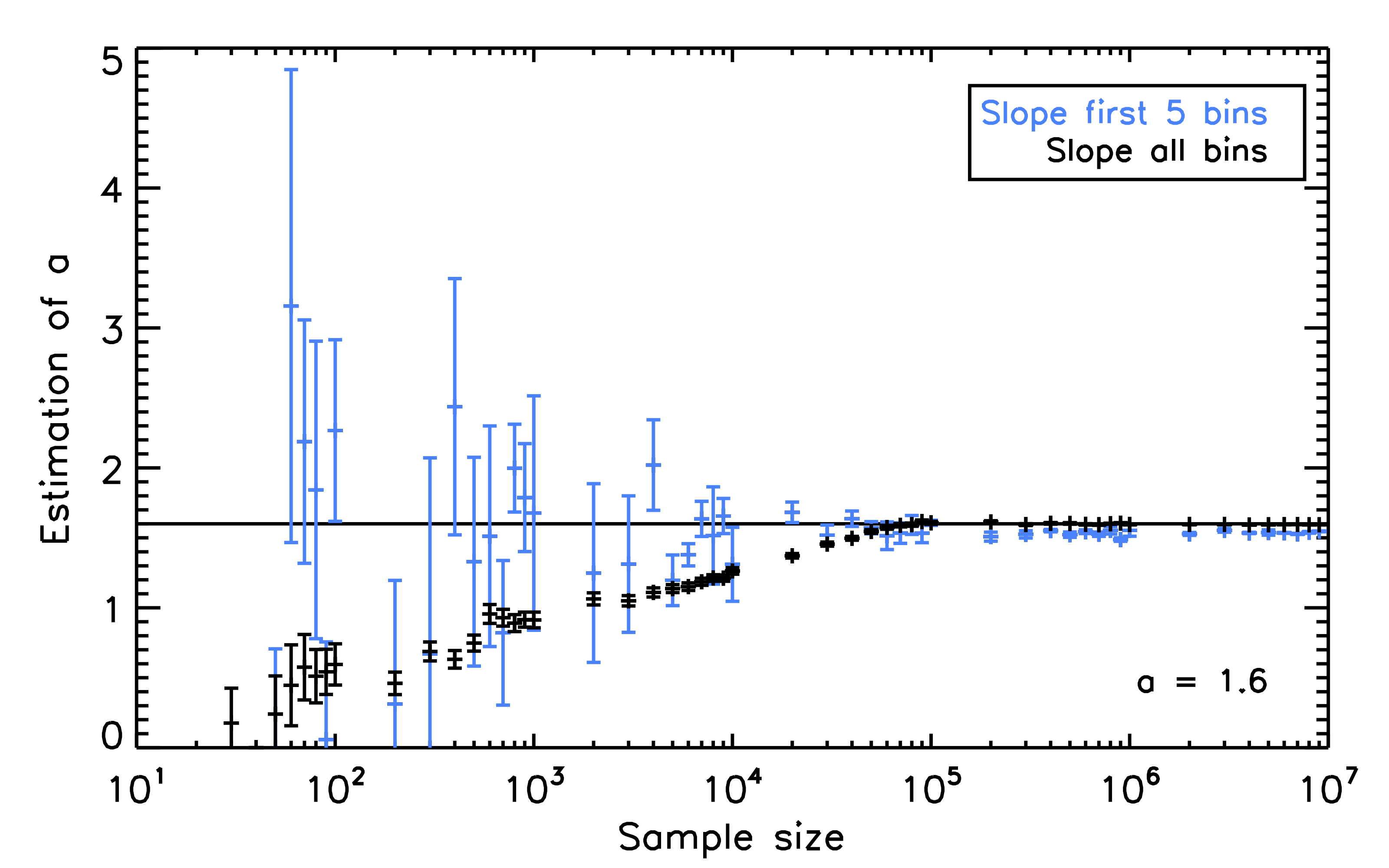}}
\caption{Evolution of the scaling parameter estimate as a function of sample size using the graphical method. The true value for $\alpha$ is $1.6$ and is indicated by the horizontal line. We plot the estimates calculated using all bins with more than one element in the sample histogram, as well as estimates for which only the first five such bins were taken into account. The error bars are defined by the $1\sigma$ uncertainty estimate for $\alpha$.}
\label{samplesizeinfluence_allbins_5bins} 
\end{figure}
 
Figure~\ref{samplesizeinfluence_allbins_5bins} illustrates the evolution of the scaling parameter estimate as a function of the sample size. Again, we used $\alpha = 1.6$ for the underlying power law. It is immediately clear that the estimates converge towards this value (indicated by the horizontal line). When all histogram bins with more than one element are used, this convergence is gradual. Conversely, for the alternative graphical method where only the first five such bins are fit and the heavy tail is ignored, the estimates are scattered around the true value of $\alpha$ with a large uncertainty.

We emphasize that a large number of data points ($n \geqslant 10^5$) is required to reliably estimate the scaling parameter graphically. This is much more than most authors realize, and for many studies this amount of data may not be available. Moreover, this required sample size increases with the size of the scaling parameter $\alpha$, as we show in the second supplementary movie. Note that for real-life datasets the true value of $\alpha$ is not known beforehand which makes it hard to assess whether a data set is sufficiently large to obtain a reliable estimate. 

\subsection{Maximum Likelihood Estimator}

\begin{figure}        
\centerline{\includegraphics[width=0.7\textwidth]{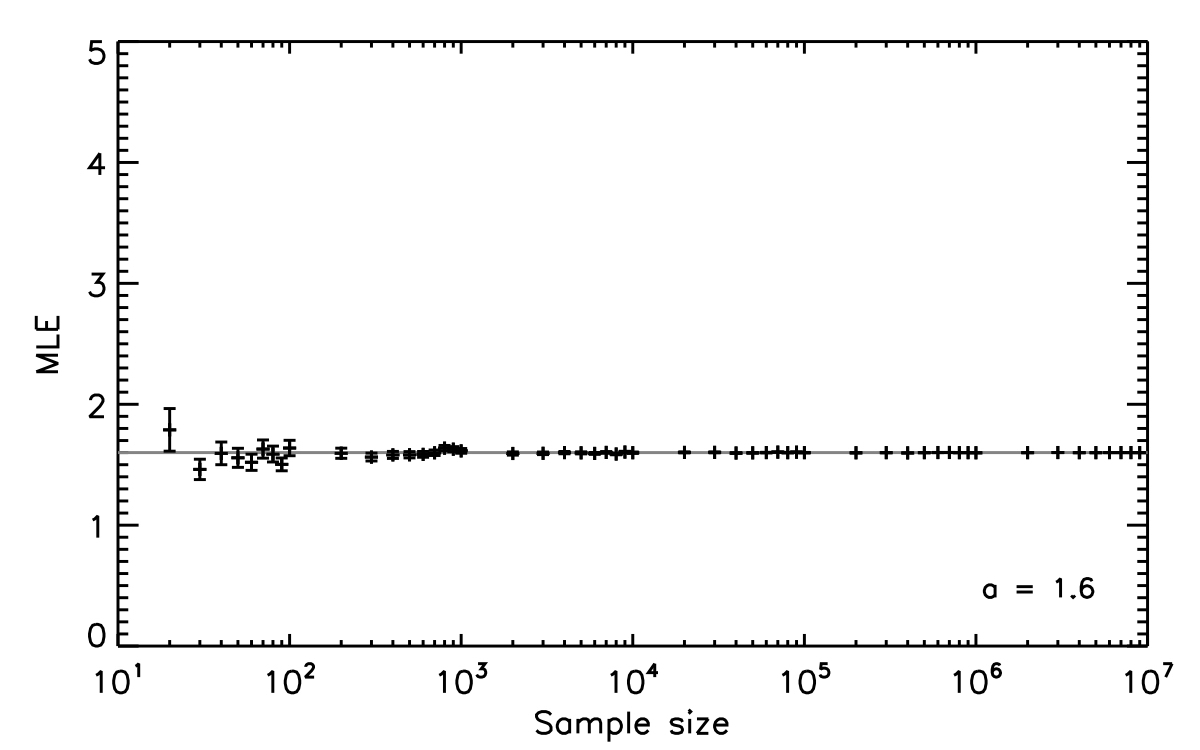}}
\caption{Evolution of the scaling parameter estimate as a function of sample size using the MLE method. The true value for $\alpha$ is $1.6$ and is indicated by the horizontal line. A reasonable estimate for $\alpha=1.6$ is obtained for a sample size as low as $n=200$.}
\label{MLEsamplesize} 
\end{figure}

The application of the MLE method to our random samples is straightforward: Equation~\eqref{MLEformula} provides the scaling parameter estimate and Equation~\eqref{MLEerrorformula} gives the standard error on this estimate. The resulting estimates for sample sizes varying between $n=20$ and $n=10^7$ are shown in Figure~\ref{MLEsamplesize}. When compared to Figure~\ref{samplesizeinfluence_allbins_5bins}, this plot illustrates the power of the MLE method very convincingly: the estimates converge rapidly to the true scaling parameter value (indicated by the horizontal line). For $\alpha=1.6$, a sample size as low as $n=200$ is sufficient to estimate this scaling parameter reliably. Note, however, that in certain studies even this sample size cannot be reached (see, for example, Section~\ref{stealthSection}).

\subsection{Method Performance}

\begin{figure}        
   \centerline{\hspace*{0.015\textwidth}
               \includegraphics[width=0.515\textwidth,clip=]{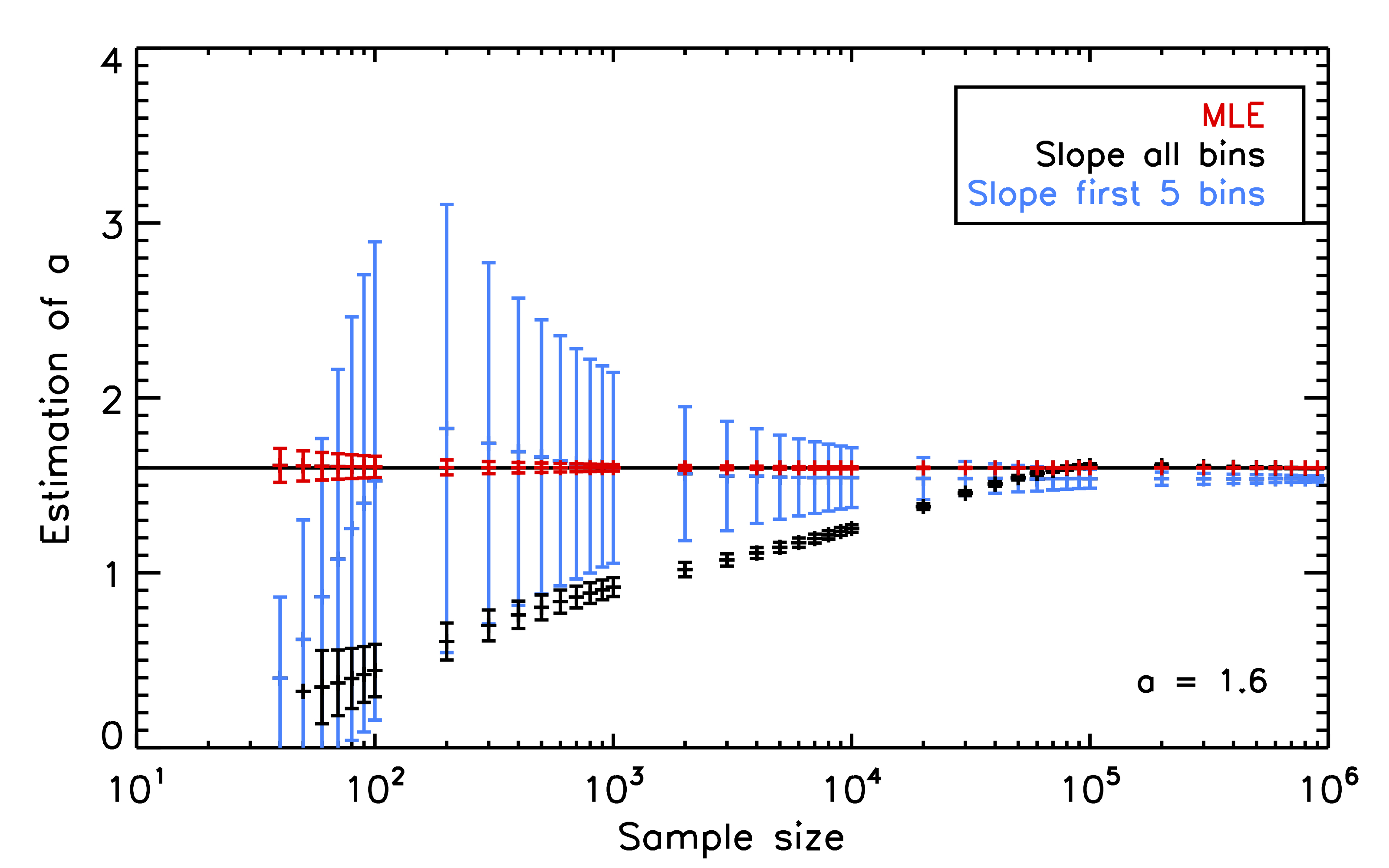}
               \hspace*{-0.03\textwidth}
               \includegraphics[width=0.515\textwidth,clip=]{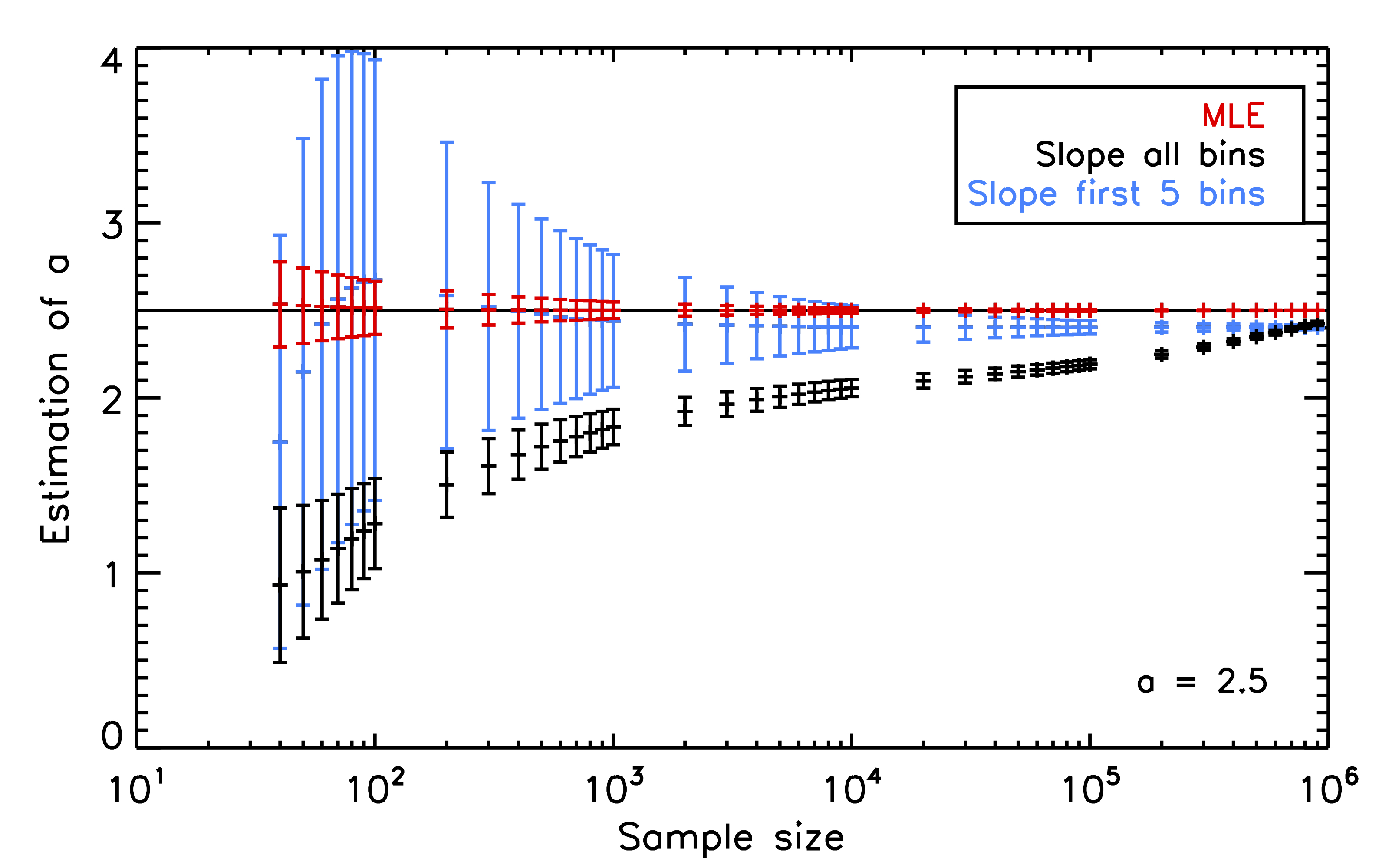}
              }
\caption{Comparison between the different estimation methods for the scaling parameter using a Monte Carlo-type simulation. Estimates (averaged over 10~000 iterations) obtained with the graphical methods are in black (all bins) and blue (first five bins). The maximum likelihood estimator is shown in red.}
\label{montecarlo_allmethods} 
\end{figure}

To compare the three methods applied above, we performed a Monte Carlo-type simulation. We generated random samples of different sizes from a power-law-distributed set and studied the average of the scaling-parameter estimates over 10~000 iterations. 

The results of this simulation for $\alpha = 1.6$ and $\alpha = 2.5$ are shown in Figure~\ref{montecarlo_allmethods}. The average estimated slope based on all histogram bins is plotted in black and converges gradually towards the true scaling parameter value. Note in the right panel, however, that even a sample size as large as $n=10^6$ is insufficient to obtain convergence for $\alpha = 2.5$. The slope estimate based on only the first five bins (in blue) converges much more quickly towards the true value. However, the actual value for $\alpha$ is never reached because the tail of the distribution is neglected, even though it becomes important as the sample size increases. Additionally, the average of the standard errors over all iterations is unacceptably large for this method. The MLE method (in red) clearly stands out as the preferred approach. The average estimates converge quickly and the standard error is low. 

In the right panel of Figure~\ref{montecarlo_allmethods}, the black estimates appear to level off prematurely in the sample size range between $n=10^3$ and $n=10^5$. If the graph were to end here, an observer could erroneously conclude that the scaling parameter estimates are converging towards the true value, while in reality the slope is steeper still, which becomes clear only when even larger sample sizes are used. 

\begin{figure}        
   \centerline{\hspace*{0.015\textwidth}
               \includegraphics[width=0.515\textwidth,clip=]{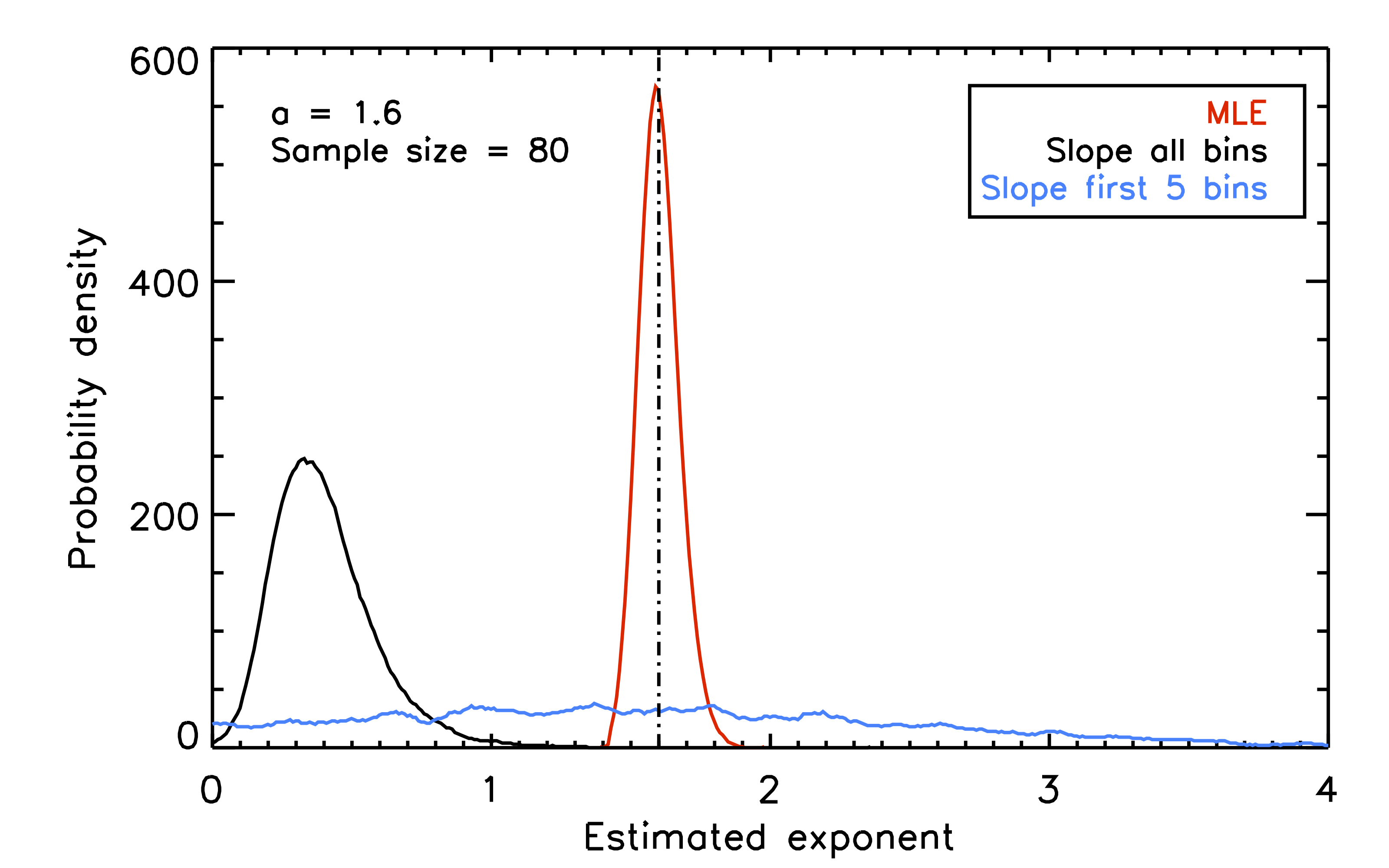}
               \hspace*{-0.03\textwidth}
               \includegraphics[width=0.515\textwidth,clip=]{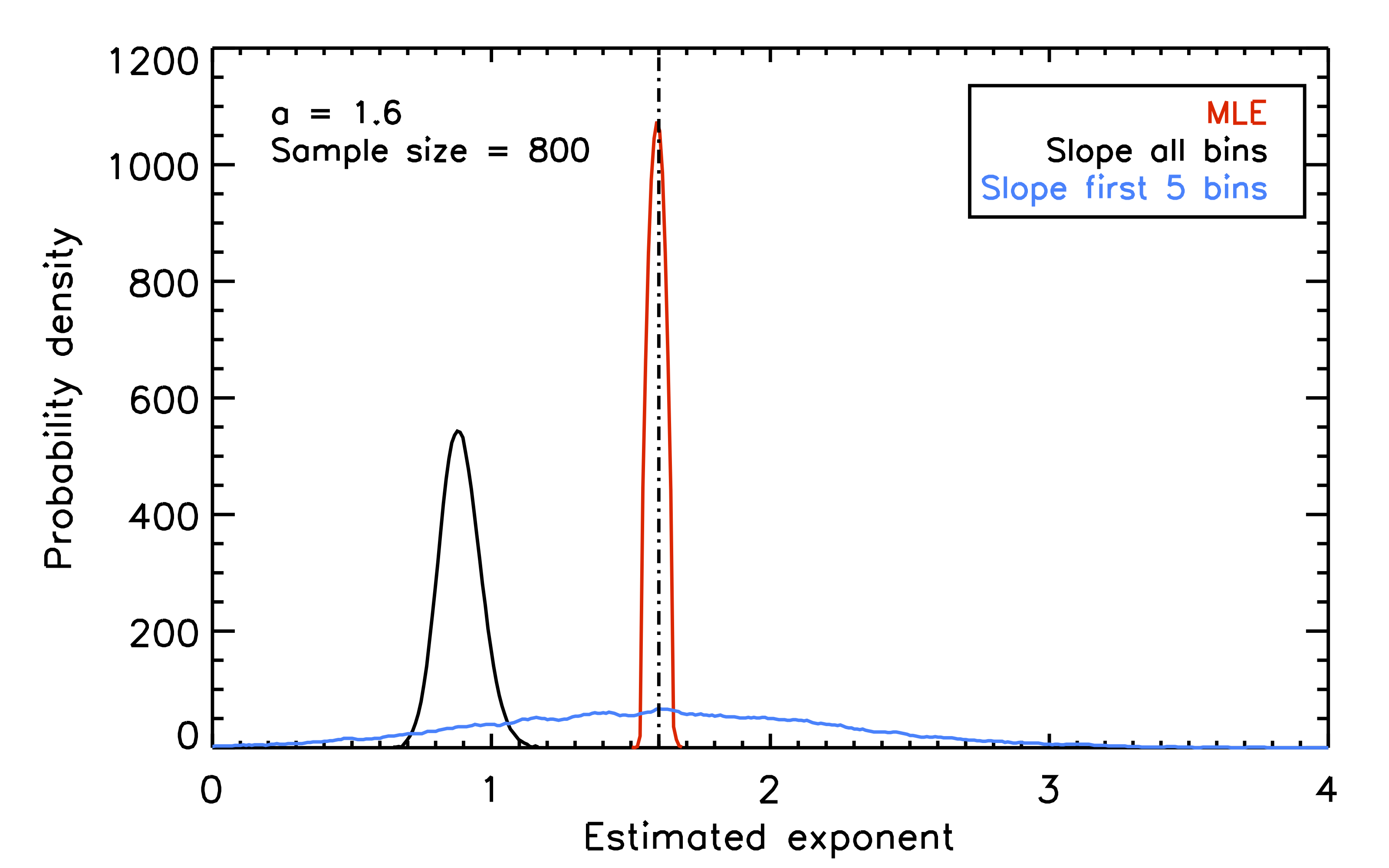}
              }
   \centerline{\hspace*{0.015\textwidth}
               \includegraphics[width=0.515\textwidth,clip=]{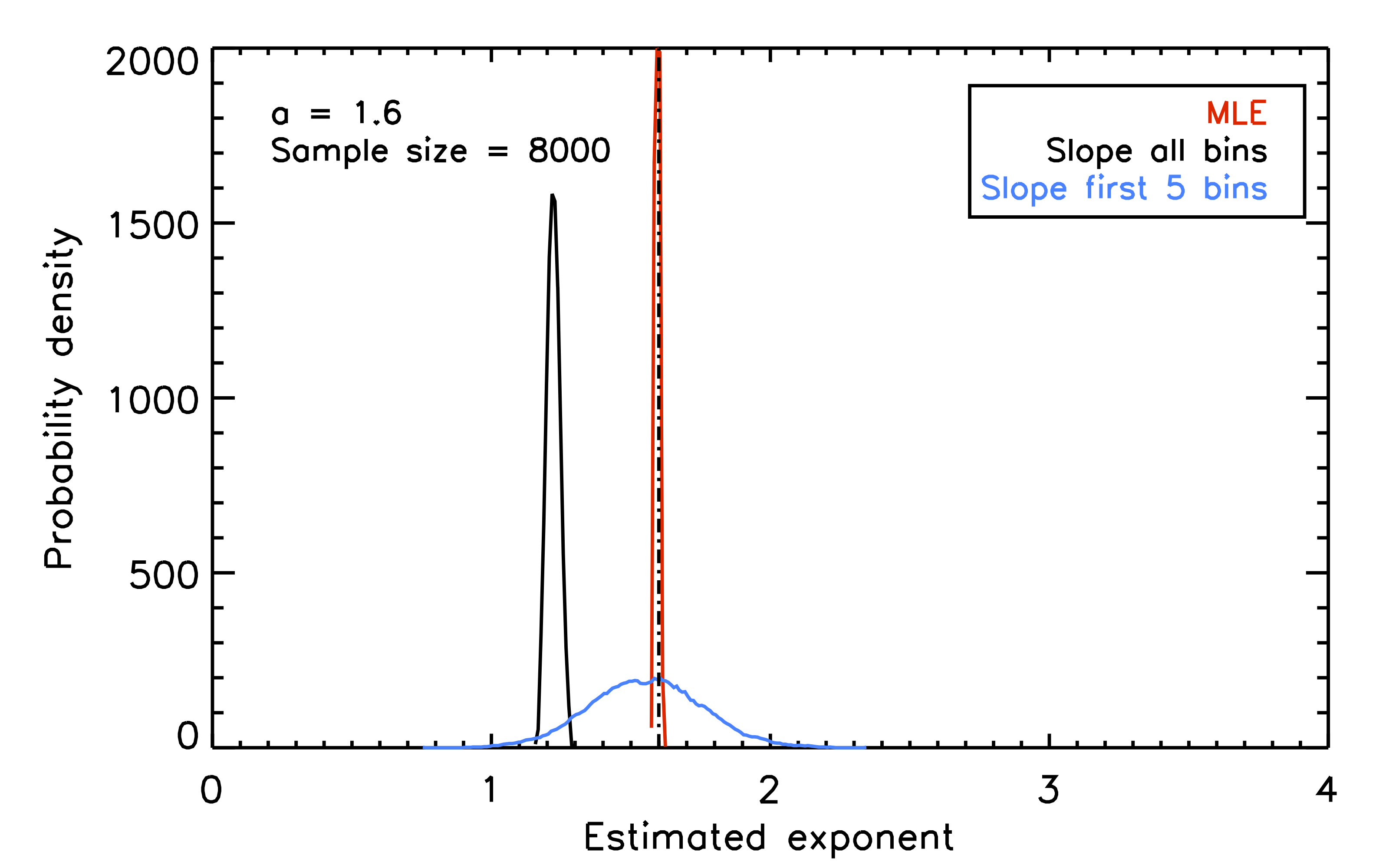}
               \hspace*{-0.03\textwidth}
               \includegraphics[width=0.515\textwidth,clip=]{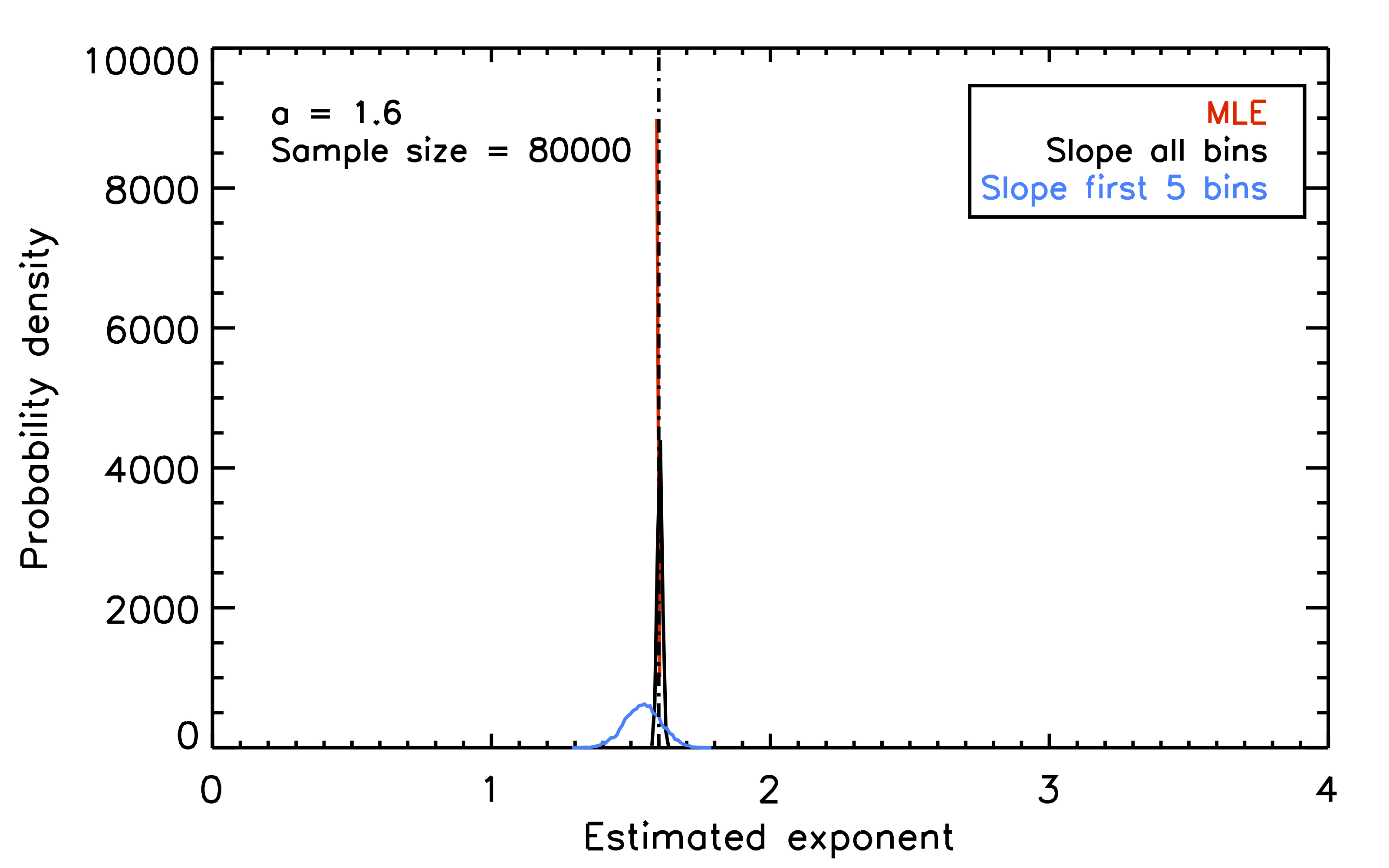}
              }
\caption{Smoothed probability density functions of the scaling parameter estimates in the Monte Carlo simulation, calculated for different sample sizes and $\alpha = 1.6$ (indicated by the vertical dotted line).}
\label{probabilitydensities} 
\end{figure}
The probability density functions for the estimates of the scaling parameter in the Monte Carlo simulation are shown in Figure~\ref{probabilitydensities} for different sample sizes and with the scaling parameter set to $\alpha = 1.6$. What stands out immediately is that for small sample sizes the spread of the estimates is very large when using the graphical method based on the first five histogram bins (blue curves). However, the density distribution encloses the true value for $\alpha$, which means that this method may by chance result in the right answer. The spread for the black curve (based on all histograms bins) is much smaller. Nevertheless, this curve is shifted to the left for the small samples, indicating that the slope is underestimated. As the sample size increases, the black curve shifts towards the true value of $\alpha$ while simultaneously reducing its spread. This results in an accurate estimation for large sample sizes. In contrast, the MLE method (red curve) performs well for all sample sizes. The probability distribution is, even in the case of a small sample, narrow and centered around the true value of $\alpha$. Better precision is achieved as the sample size increases, and the probability distribution narrows even further. 

\begin{figure}        
\centering
\includegraphics[width=0.7\textwidth]{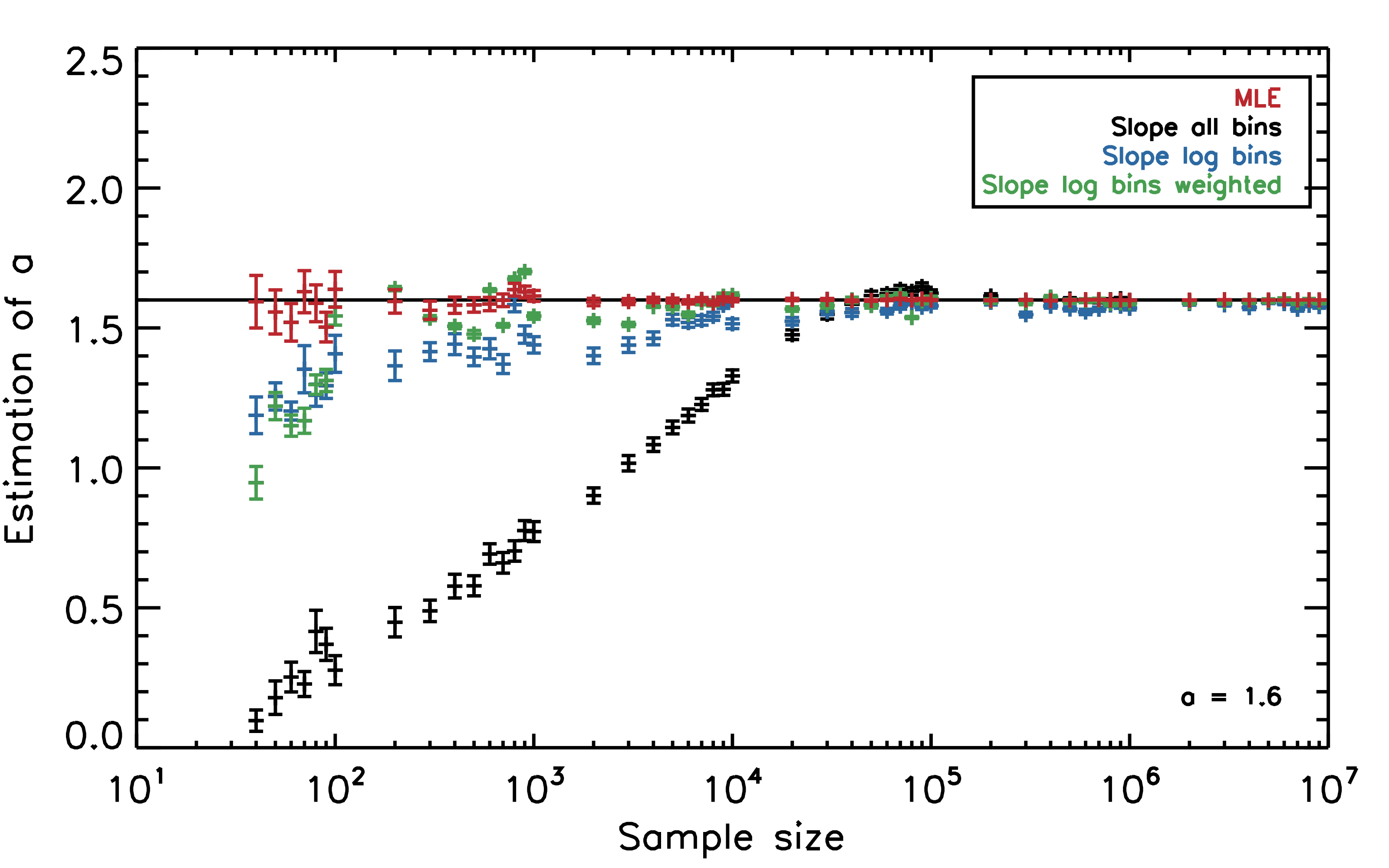}
\caption{Evolution of the scaling parameter estimate as a function of sample size using the improved graphical methods. The true value for $\alpha$ is $1.6$ and is indicated by the horizontal line. We show the estimates for $\alpha$ based on linear binning (in black), logarithmic binning (in blue), logarithmic binning with weights (in green), and the MLE method (in red).}
\label{logbinweights} 
\end{figure}

\subsection{Improvements to the Graphical Methods}
\label{logbinsection}
Several modifications to the graphical methods are possible to improve their performance. The binning method has a large influence on the estimates obtained for the scaling parameter. When the histogram is binned linearly, the estimates for small sample sizes are far below the true value. Applying a least-squares method to logarithmically binned histograms for the same samples improves the estimates considerably, as is clear from the blue points in Figure~\ref{logbinweights}, where we applied a logarithmic bin size of 0.1 and normalized according to the linear bin width. Figure~\ref{logbinweights} also shows that while the graphical method performs better with logarithmic bins, it remains unreliable for small sample sizes. 

Using weights for the linear regression can improve the estimates even further. We have applied a weight $w$ to each bin according to $n$, the number of data points in that bin: $w = \sqrt n$. These weights correspond to the uncertainty in each bin and result from binomial counting statistics. In case of a sufficiently large sample, the binomial distribution can be approximated by a normal one with standard deviation $\sigma=\sqrt{Np(1-p)}$, where $p$ is the probability of being in the bin and $N$ is the total number of data points in all bins. The probability $p$ can then be approximated as $n/N$, giving the uncertainty estimate $\sigma = \sqrt{n(1-n/N)} \approx\sqrt n $ for $n \ll N$. 

The green estimates in Figure~\ref{logbinweights} illustrate the additional improvement that this weighing of the bins brings. Nevertheless, the MLE results (in red) outperform also these improved graphical estimates. 

Many authors have indeed applied logarithmic (and sometimes even weighted) binning in earlier studies and their estimates are thus more reliable than suggested by the worst-case scenario of the linearly binned graphical methods that were shown before. Nevertheless, the resulting estimates will still prove less accurate than the ones obtained with MLE, especially in case the samples used in these studies were small in size. We therefore encourage authors to re-analyze their data using this MLE method and to treat non-MLE results, even the ones obtained with logarithmic and weighted binning, with caution. (Note that we have presented a highly idealised scenario here using perfectly power-law-distributed samples. For real-life datasets that contain uncertainties, a careful fitting of the logarithmically binned histogram may yield quite robust results as well.)

\section{Application to Solar Data}

This study led us to the realization that we need to revisit some of our own work where we have analyzed power laws in solar data. Recently, we investigated a scaling law for the angular width of coronal mass ejections \citep{DHuys2014}; while one of us \citep{Berghmans1998} studied the power-law-distributed radiative losses of quiet-Sun coronal brightenings. In both studies, we applied the graphical method to a log-transformed, linearly-binned histogram of our dataset. The discussion above clearly illustrates that this is not the preferred approach. We therefore repeat our earlier analysis, now applying the MLE method to our data, and discuss the differences in the results.

\subsection{CMEs without Low Coronal Signatures}
\label{stealthSection}

We investigated coronal mass ejections (CMEs) without low coronal signatures (LCS): solar eruptions, detected as CMEs in coronagraph observations, for which no on-disk signatures such as flares or filament eruptions are observed \citep{DHuys2014}. These events are commonly referred to as \textit{stealth CMEs}. We studied their observational characteristics and asked how they differ from classical CMEs. One characteristic we studied in detail is the scale invariance of CME angular widths. This property was discussed by \cite{Robbrecht2009} in their analysis of the CACTus LASCO CME catalog (\url{www.sidc.be/cactus}). These authors found a power law with an average scaling parameter $\alpha~\approx~1.66$ over 10 years of CME detections (1997 to 2006, based on the LASCO level zero images). We repeated this analysis for the case of CMEs with and without LCS, observed in 2012, and found a notably different slope for both classes of events: $\alpha~=~0.97~\pm~0.41$ for CMEs without LCS, and $\alpha~=~1.50~\pm~0.34$ for classical CMEs (based on the LASCO quick-look images). 

During that study, we noticed the important influence of the sample size on the derived slope. When selecting only a small sample of classical CMEs, the slope resulting from the graphical method was notably flatter than the slope derived for the complete set of CMEs in 2012. The observation of the effect of the sample size on our stealth CME results was an important motivation for the work presented in the current paper.

Our research here suggests that applying the MLE method to the CME data set should yield far more reliable results. First, we need to select a suitable value for the lower cutoff, $x_{\rm min}$. As explained before, we choose the value for $x_{\rm min}$ that minimizes the difference between the probability functions for the empirical data and for the best-fit power law above this $x_{\rm min}$. The distance between these two probability functions is calculated by the Kolmogorov-Smirnoff (KS) statistic: $D = \max_{x \geqslant x_{\rm min}} |S(x)-P(x)|$.
In this equation, $P(x)$ is the cumulative distribution function (CDF) for the best-fitting power law, and is given by the following formula \citep{Clauset2007}:
\begin{equation}
P(x) = \left( \frac{x}{x_{\rm min}}\right)^{-\alpha+1}.
\end{equation}
$S(x)$ denotes the CDF  for the observations, which describes the probability that a variable \textit{X}, randomly drawn from the data, will have a value $X \geq x$. Therefore $S(x)$ can be written as the vector $[1, \frac{n-1}{n}, ..., \frac{2}{n}, \frac{1}{n}]$ \citep{Lai}. We can then calculate the KS statistic for each possible value of $x_{\rm min}$. The preferred value for this parameter is the one that minimizes \textit{D}.

We implemented this procedure for the observations of the angular width of stealth and normal CMEs. The lowest value for $x_{\rm min}$ we considered was $x_{\rm min} = 5\degree$, as this is the lower detection limit set by the CACTus program. For CMEs without LCS we used an upper limit of $x_{\rm min} = 55\degree$ in order to retain at least five data points to construct the probability functions with. For the regular CMEs, the upper limit was set to $x_{\rm min} = 120\degree$, the maximal angular width for non-halo CMEs, as set by CACTus. Indeed, we exclude halo CMEs because their angular width is not well-defined due to strong projection effects in the observations. The resulting estimate for both classes of CMEs was $x_{\rm min} = 5\degree$ and we will therefore use this parameter value for the remainder of the discussion. 

Our study of CMEs without LCS was motivated by the question of whether and how this class of events differs from classical CMEs. With $x_{\rm min}=5\degree$, we find that the MLE value for the stealth CMEs and the classical CMEs are, respectively, $\alpha=1.735~\pm~0.116$ and $\alpha=1.657~\pm~0.017$, indicating that the MLE values for both classes of CMEs are compatible within the error margins. This suggests that CMEs without LCS may not be so different from events that do show these features, at least as far as their angular width distribution is concerned. Indeed, no difference between both CME classes can be derived from these power-law estimates.

Note that the small number of stealth CMEs (we found only 40 events) means that even the MLE value cannot be trusted completely, although it should be more reliable than the slope we found earlier using a linearly binned histogram. The group of classical CMEs contained nearly 1600 events, which should be sufficient to obtain a reliable MLE value (as is also clear from the small error margins). In contrast, for graphical methods this sample size is too small. Clearly, the slopes and error margins we determined earlier differ significantly from the MLE results due to the limited sample sizes. However, we are drawing conclusions based on a very small sample of stealth CMEs. A thorough answer to the question of whether stealth CMEs are physically different from classical CMEs really requires more data. Unfortunately, these stealth CMEs are not easily identified, which makes expanding our data set challenging and very time-consuming.

Finally, we can compare the yearly maximum likelihood estimator based on the CME widths in the CACTus LASCO quick-look catalog to the scaling parameters obtained by \cite{Robbrecht2009} for the years 1997 to 2006. These values are shown in Table~\ref{MLEyears} for both time periods. Interestingly, the MLE estimates for the years 2010 to 2014 lie very close together, which increases our confidence in the values we obtained. 

Though the results reported by \cite{Robbrecht2009} are mostly in agreement with ours, there is quite some spread in their estimates. It is likely that, for the deviating years, they suffered from the limitations of the graphical method they applied, especially in light of the fact that, particularly at the time of solar minimum, the number of CMEs per year was presumably quite small. Another influence on their estimate may come from unusual solar events. In 2003, for example, the very strong Halloween storms were observed. Atypical CMEs like these may have skewed the statistics. Also here the influence of one unusual event on the estimate is especially large in case of small sample sizes.

For the MLE calculations applied to the years 2010, 2011, 2012, 2013, and 2014 samples of, respectively, 221, 1057, 1289, 1381, and 1470 CMEs were used. These sample sizes should be sufficient to obtain a reliable MLE result, but not a graphical one. To increase our sample size even more, we calculated the MLE for all CMEs observed in the period covering from July 2010 to December 2014 (spanning the complete CACTus LASCO quick-look catalog) and obtained a scaling parameter estimate $\alpha = 1.677 \pm 0.008$. This MLE value is close to $\alpha~\approx~1.66$, the average value over all years found by \cite{Robbrecht2009}. The strong correspondence between the MLE values over the years, the MLE value based on all data, and the average slope value strongly suggest that solar eruptions, regardless of their size or when they occur during the solar cycle, are initiated by the same physical mechanism.

\begin{table}
\caption{MLE and slope estimations for CMEs detected by CACTus for various years. Slope parameters for the years 1997 to 2006 were calculated using a graphical method by \cite{Robbrecht2009}.}
\label{MLEyears}
\begin{tabular}{ccccc}      
\hline         
Year & Slope & & Year & MLE \\
  \hline
1997 & $\alpha = 1.49$ & & 2010 & $ 1.68 \pm 0.04 $ \\
1998 & $\alpha = 1.64$ & & 2011 & $ 1.70 \pm 0.02 $ \\
1999 & $\alpha = 1.68$ & & 2012 & $ 1.66 \pm 0.02 $ \\
2000 & $\alpha = 1.64$ & & 2013 & $ 1.68 \pm 0.02 $ \\
2001 & $\alpha = 1.63$ & & 2014 & $ 1.67 \pm 0.02 $ \\
2002 & $\alpha = 1.67$ & & & \\
2003 & $\alpha = 1.97$ & & & \\
2004 & $\alpha = 1.83$ & & & \\
2005 & $\alpha = 1.71$ & & & \\
2006 & $\alpha = 2.04$ & & & \\
 \hline
\end{tabular}
\end{table}

\subsection{Quiet-Sun EUV Brightenings}
\cite{Berghmans1998} studied the radiative losses of transient brightenings in the quiet-Sun regions using the {\it Extreme Ultraviolet Imaging Telescope} (EIT) 304~\AA \ images from the {\it Solar and Heliospheric Observatory} \citep[SOHO;][]{Delaboudiniere1995}. The study of the energy distribution of solar brightenings has been a popular subject ever since \cite{Hudson1991} proposed that a power-law exponent $\alpha=2$ is the critical value deciding whether coronal heating can be explained by nanoflares. If the small events outnumber the larger ones sufficiently (corresponding to a steep power law: $\alpha > 2$), these nanoflares would contain enough energy to heat the solar corona, while remaining below the detection limit. \cite{Berghmans1998} reported a scaling parameter $\alpha = 1.9 \pm 0.1$ for their measurements of the radiative losses of impulsive EUV brightenings in the transition region, which suggests that the majority of the energy is provided by the larger events, although the difference between this value and the critical slope value is marginal. 

We revisited the original data set observed by SOHO/EIT on 28 December 1996 and reconstructed the event detection as described by \cite{Berghmans1998}. We did not apply any pre-processing (flat-field, grid pattern correction, solar rotation compensation, {\it etc}.) as \cite{Berghmans1998} applied these in a non-standard way ({\it i.e.} they did not use the \textsf{eit\_prep} solar software) and this preprocessing has in any case no real effect on the statistics of event detection. More important to note is that we also chose to skip removal of cosmic ray hits as these cannot be reliably distinguished from the smallest brightenings in the solar atmosphere.

For the event detection algorithm itself we closely followed the method described in Section 4.1 of \cite{Berghmans1998}, thereby using a running-average of 1 h to produce the reference background for each pixel. We used the $\Sigma_P=2.5$ and  $\Sigma_E=1.5$ thresholds for the peaks and their intensity to obtain 133~262 brightening detections. Figure~\ref{HistogramNanoflares} shows a linearly binned histogram of the radiative losses of these events, as well as a logarithmically binned one\footnote{Our dataset expresses energy in data numbers (DN). \cite{Berghmans1998} report a factor of $2\times10^{20}$ erg per DN to convert the flare energies to the physical units used in the histograms. However, we suspect a typographical error in this conversion factor. We needed to rescale our histogram by a factor of $2\times10^{24}$ erg per DN to reproduce the results of \citeauthor{Berghmans1998} and to obtain reasonable energies for solar flares.}. Our new analysis finds an order of magnitude more events compared to the 13~067 events found by \cite{Berghmans1998}. Note however that in contrast to Figure~17 of \cite{Berghmans1998}, our new linearly binned histogram in Figure~\ref{HistogramNanoflares} does not depart from a power law for small events. We thus attribute the difference in the number of events to the class of the smallest scale brightenings which are easily confused with cosmic ray hits.

\begin{figure}
   \centerline{\hspace*{0.015\textwidth}
               \includegraphics[width=0.515\textwidth,clip=]{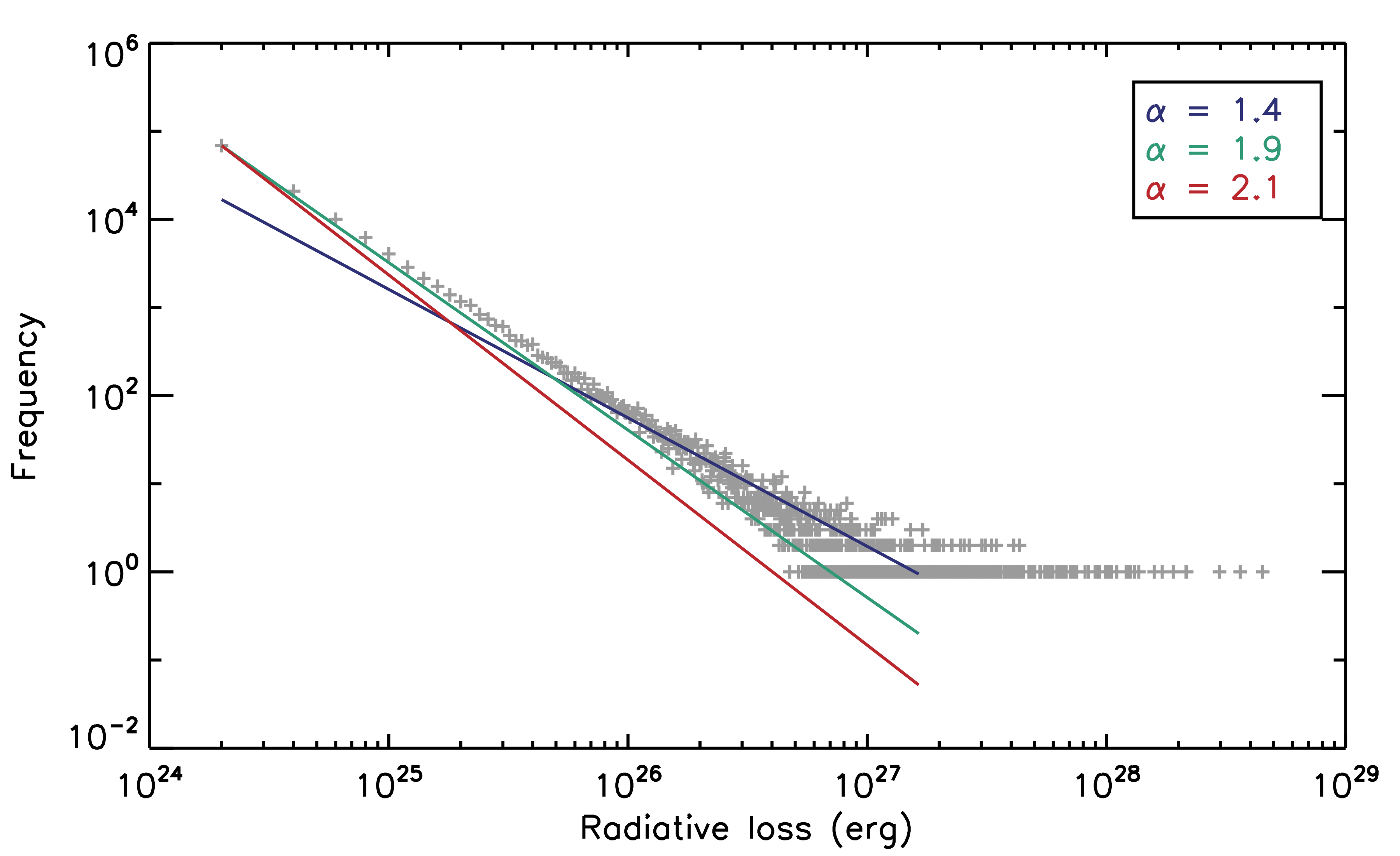}
               \hspace*{-0.03\textwidth}
               \includegraphics[width=0.515\textwidth,clip=]{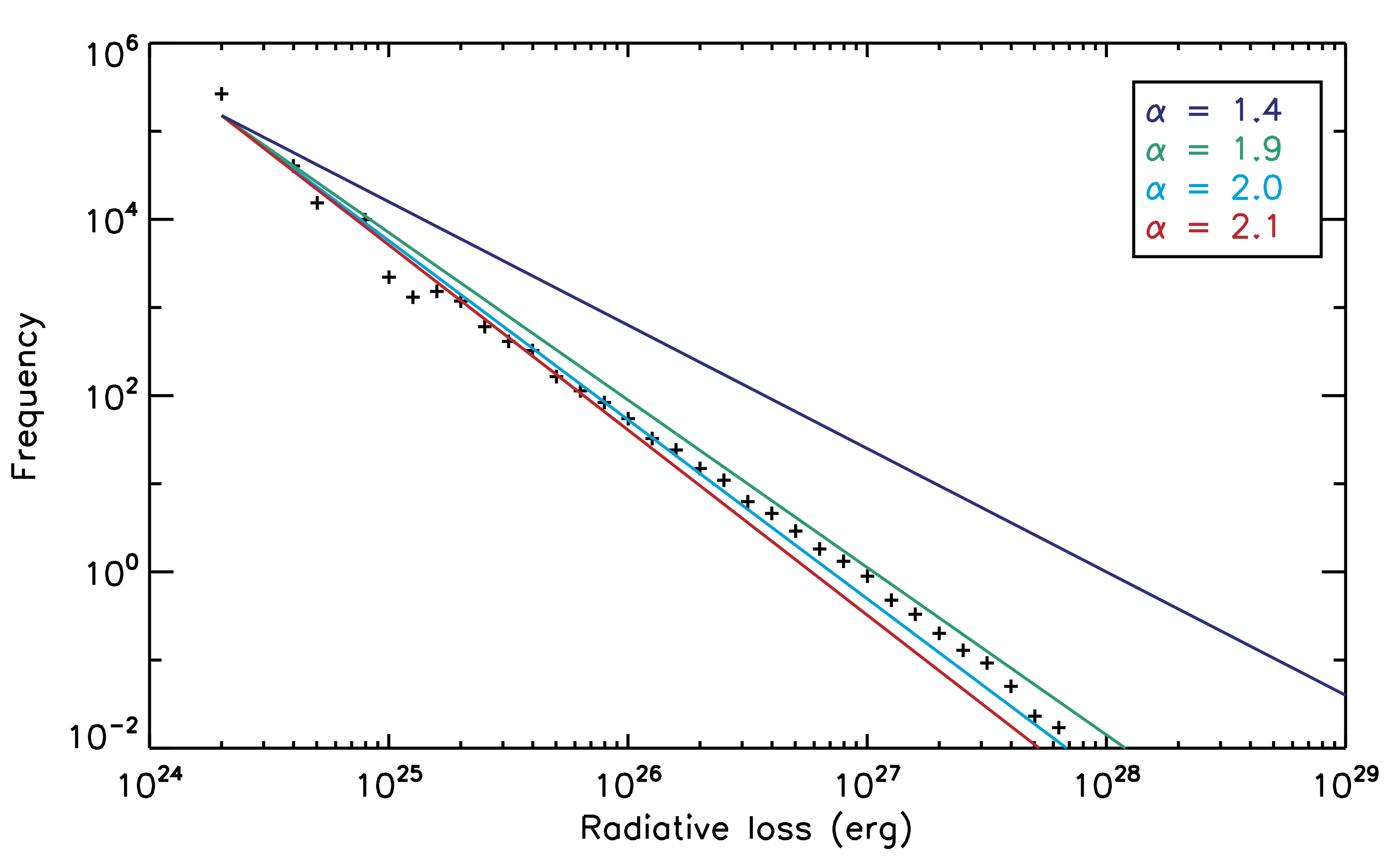}
              }         
\caption{Left: Frequency histogram for the nanoflare radiative loss, binned linearly.  Overplotted are lines with slope $\alpha = 1.9$ \cite[as reported by][in green]{Berghmans1998}, $\alpha = 1.4$ (the result of a linear fit on the entire data sample, in blue) and $\alpha = 2.1$ (MLE, in red). The slope for the MLE value and the value found by \citeauthor{Berghmans1998} are anchored in the first data point of the histogram. Right: Logarithmically binned frequency histogram for the radiative losses of nanoflares. The same slopes are overplotted, with the addition of  $\alpha = 2.0$, the result of a linear fit on the logarithmically binned histogram (light blue). The slopes are all anchored in the starting point of the light blue fit to the logarithmic histogram (with $\alpha = 2.0$). }
\label{HistogramNanoflares} 
\end{figure}

We applied a linear fit to the linearly binned histogram of the radiative losses of all detected brightenings. This graphical method yielded a slope value of $\alpha = 1.45$, lower than that reported by \cite{Berghmans1998}. However, the left panel of Figure~\ref{HistogramNanoflares} clearly shows that the linear fit is strongly influenced by the heavy tail of the histogram and that a slope value of $\alpha = 1.9$ is a good fit to the linear part of the histogram. 

As we have shown in Section~\ref{logbinsection}, logarithmic binning improves the estimate of $\alpha$. In the right panel of Figure~\ref{HistogramNanoflares}, we used a logarithmic bin width of $0.1$ and normalized by dividing by the bin width. Figure~\ref{HistogramNanoflares} clearly shows that this type of histogram solves the heavy tail problem by combining more data points in the larger bins at the high end. This results in better estimates for $\alpha$, provided that the sample size is large enough. A linear fit to the logarithmically binned histogram gives us a slope value of $\alpha = 2.0$. 

However, applying the MLE method to these data paints a different picture still. We obtain a scaling parameter value $\alpha = 2.10 \pm 0.003$, which is above the critical value proposed by \cite{Hudson1991}. This value was obtained by setting the parameter $x_{\rm min} = 4.66\times10^{26}$, as determined using the KS statistic. As an extra condition, we made sure that at least $1000$ data points remained for the MLE calculation. Based on Figure~\ref{montecarlo_allmethods}, we consider this a sufficient number to estimate a scaling parameter with a value around $2$ reliably. Note that this high value for $x_{\rm min}$ means that, for this data set, the power law does not hold for lower energies and no conclusion on nanoflare heating can be drawn, even though our MLE estimate is above the critical $\alpha=2$ value. This is exactly the caveat pointed out by \cite{Berghmans2002}: the steepness of the power law is a necessary, but not a sufficient condition for coronal heating by nanoflares. The diffferent behavior for lower energies may be physical, or it may be a result of the applied detection technique. It is conceivable, for example, that it is a consequence of the possible confusion between very small brightenings and cosmic rays.

We emphasize once more that the sample size here is of the order of $n~\approx~10^5$ elements, which may seem sufficient to many authors, even when applying graphical methods. Figure~\ref{montecarlo_allmethods} (right panel) suggests, however, that for a true value of $\alpha = 2.1$ there is a levelling in the slope estimates when $10^3 \leq n \leq 10^5$, which ---incorrectly--- suggest convergence is reached. An even larger sample size is needed to reliably estimate the scaling parameter using a graphical method like the one applied by \cite{Berghmans1998}. We assume that had the authors used a more robust estimation method for the scaling parameter, for example the MLE method, the conclusions of their work would have differed. 

Note also that \cite{Ryan2016} explored the influence of various thresholds used in flare detection algorithms on the power-law parameter estimate for the resulting distributions of the flare peak flux. Applying MLE, they find that the power-law exponents of these distributions are not stable but appear to steepen with increasing peak flux. This deviation from a true power law may again be the result of the detection and analysis techniques that are applied. However, it could also imply that the observed flare size distribution is not a true power law. Instead, this distribution may be a convolution of multiple heavy-tailed distributions, resulting, for example, from different active regions or different flaring mechanisms. 

\section{Conclusions}

Power-law distributions are found in various solar data sets and the accurate estimation of their scaling parameter is of great interest. For instance, these values are used to validate theoretical predictions. If small and large scale events turn out to exhibit the same power-law behavior, this may also point at a common physical process causing them. Therefore, it is vital to estimate this scaling parameter as accurately as possible.

Many  authors continue to use graphical methods to obtain an estimate of the scaling parameter when they believe they have observed a power-law behavior in their data. While many of them may have found a reliable power law, some of these results need to be revisited. Indeed, we have shown that these graphical methods often result in inaccurate estimates, especially in cases with small sample sizes. In fact, the sample sizes required in order to obtain reliable results with a graphical method are so large that they may not be achievable in certain studies. In addition to the problem of small sample sizes, the graphical estimate of the scaling parameter depends strongly on the binning method, which makes the result even less trustworthy.

In view of our analysis of existing studies of power laws in solar physics we strongly recommend that authors who have published scaling parameter estimates based on a graphical method review and reconsider their analyses using more appropriate techniques, such as MLE or even more general methods (for example a Bayesian approach). Nevertheless, the problem of sample size remains for these techniques as well: a sufficiently large sample is needed to obtain a reliable estimate. Even though this minimally required number of data points is vastly lower for the MLE method compared to the graphical methods, it may still be too large for some studies. In that case, we would advise authors to be extremely prudent in drawing conclusions based on any estimation, even those made using the MLE method. Likewise, previous studies in which a graphical method --- or even the MLE method --- was applied to a limited amount of data should be treated carefully.

\begin{acks}
The authors are grateful to V. Delouille and the PROBA2/SWAP team for valuable input. We also thank the anonymous referee for insightful comments that helped us improve this paper. This research was co-funded by a Supplementary Researchers Grant offered by the Belgian Science Policy Office (BELSPO) in the framework of the Scientific Exploitation of PROBA2, the Inter-University Attraction Poles Programme initiated by BELSPO (IAP P7/08 CHARM), and the European Union's Seventh Framework Programme for Research, Technological Development and Demonstration under Grant Agreements No~284461 (Project eHeroes, www.eheroes.eu) and No~269299 (Project SOLSPANET, www.solspanet.eu). These results were also obtained in the framework of the projects GOA/2015-014 (KU Leuven), G.0729.11 (FWO-Vlaanderen) and C~90347 (ESA Prodex). E.~D'Huys and D.B.~Seaton additionally acknowledge support from BELSPO through the ESA-PRODEX program, grant No.~4000103240. This paper uses data from the CACTus CME catalog, generated and maintained by the SIDC at the Royal Observatory of Belgium (www.sidc.be/cactus). 
\end{acks}
\fi
\bibliographystyle{spr-mp-sola}
\bibliography{powerlaw_bibliography} 

\end{article}
\end{document}